# TOA Positioning for a TDMA Localization System

Sihao Zhao, Xiaowei Cui, Shuang Xu, Mingquan Lu
Department of Electronic Engineering
Tsinghua University
Beijing, China

*Abstract*— Positioning with one single communication between base stations and user devices can effectively save air time and thus expand the user volume to infinite. However, this usually demands accurate synchronization between base stations. Wireless synchronization between base stations can simplify the deployment of the positioning system but requires accurate clock offset estimation between base stations. A time division multiple access (TDMA) localization system in which user devices only receive signals from base stations to generate time of arrival (TOA) measurements to position themselves and no cables are needed to interconnect base stations for clock synchronization is proposed, implemented and tested. In this system, the user devices can easily join in or exit without influence to other users and the update rate of each user can be easily adjusted independently according to its specific requirement.

## I. INTRODUCTION

A typical radio frequency positioning system is usually comprised of infrastructures and user devices between which one-time or more communications are conducted to obtain the user position[1]. With more than one communications, the distances between the infrastructures and the user device can be directly obtained. This scheme is easy to implement but the user capacity and position update rate for each user are limited due to the air time occupation[2]. Using one-way communication can alleviate or even overcome this problem. For example, if the infrastructures transmit signals and the user devices only receive them, the system can achieve an unlimited user volume. Global navigation satellite systems (GNSS) are representative systems out of this idea, in which multiple infrastructures - the navigation satellites, simultaneously transmit signals using frequency division multiple access or code division multiple access[3-5].

In this work, a positioning system with broadcasting infrastructures and user receivers is designed and implemented. The infrastructures or base stations uses a time division multiple access (TDMA) to broadcast their positioning signals. These signals are received by the receiver to calculate the user position and also received by base stations to estimate the time offsets between them.

Firstly, the time of arrival (TOA) measurement model is given to illustrate how the time offsets between base stations are computed. The relationship between the user position and the pseudorange measurements from different reception epochs is also established and analyzed. A positioning approach using these pseudoranges from different epochs and different base stations is proposed. Then the proposed scheme is implemented using ultra-wideband transmitters and receivers to form a positioning system. With this system, tests are conducted to verify the performance of the proposed algorithm. Results show that the proposed positioning algorithm can calculate the user positioning correctly.

With such a TDMA scheme, the base stations can estimate their time offsets and broadcast positioning signals to the receiver at the same time, which saves the air time, removes cables for synchronization, decouples the connection between the base stations and the user devices and makes the user volume unlimited. For each user device, its position update rate can be easily adjusted and the upper limit is only associated with the broadcast frequency of the base stations and the computational capacity of its own, which makes the user device both electromagnetic silent and flexible. Then, tests are conducted to validate the proposed system scheme and the positioning approach.

## II. TRANSMISSION AND RECEPTION MATHEMATICAL MODEL

In order to achieve a positioning system with unlimited user volume, it is a natural idea to make it a broadcasting system in which base stations broadcast signals and the user devices receive and process the signals to obtain their own positions. However, due to limited spectrum resources, a code division multiple access (CDMA) or TDMA implementation can be adopted. In this work, a TDMA scheme is used to avoid pseudorandom code design and for simplicity of system implementation.

A simple signal reception model is presented as a foundation of the transmission and reception mathematical model for our TDMA positioning system as given in (1).

$$T_{irx} = T_{jtx} + b_{ji}(T_{irx}) + D_{ij} + d_{jtx} + d_{irx} \qquad (1)$$

where $T_{jtx}$ is the transmit time of the *j*th transmitter or base station measured locally in our system, $T_{irx}$ is the local reception time of the *i*th receiver which can either be a user device or another base station, $b_{ji}(T_{irx})$ represents the clock offset between the transmitter and the receiver, i.e. the time of receiver *i* minus that of transmitter *j*, at the instant of $T_{irx}$, $D_{ij}$ is the travel time of light from the *j*th transmitter to the *i*th receiver, $d_{jtx}$ and $d_{irx}$ are constant transmitter and receiver delays, respectively. All the variables have an identical unit of second in this and the following equations.

Specifically, for a user device, i.e. a receiver, the so-called pseudorange measurement can be modeled as (2).

$$\rho_i(t_i) = t_i - T_{itx} = D_i(t_i) + b_i(t_i) + d_{itx} + d_{rx} + \varepsilon(t_i) \quad (2)$$

where $t_i$ is the reception time, $T_{itx}$ is the transmission time from the *i*th base station, $D_i(t_i)$ represents the distance between the receiver and the *i*th base station, $b_i(t_i)$ is the clock bias between the receiver and the base station, i.e. the time of the receiver minus that of the base station, $d_{itx}$ is the constant transmission delay, $d_{rx}$ is the receiver processing delay and $\varepsilon$ is the measurement noise.

(2) is very similar to the pseudorange equations used in GNSS receivers. However, there are still some issues to be resolved. The first one is how to synchronize all base stations or obtain the clock relationships between them. One simple idea is to use cables to connect all base stations to transfer clock pulses to synchronize them. Obviously, this is not convenient and cost-effective for system implementation and installation. A wireless "synchronization" can be possibly achieved if the actual clock offsets between base stations can be computed. Provided this synchronization information, the clock offsets between the receiver and all base stations, i.e. $b$ in (2), can be represented by the clock offset between the receiver and one specific base station, e.g. the *i*th base station, as given by (3).

$$b_j(t_j) = b_i(t_j) - b_{ij}(t_j) \quad (3)$$

It should be noted that, in (3), the clock offset between the receiver and the *j*th base station at time $t_j$ can be represented by the clock offset between the receiver and the *i*th station at the same instant. However, due to the clock drift, the clock offset $b_i$ at $t_j$ is different from that at $t_i$, and their relationship is modeled by .

$$b_i(t_j) = b_i(t_i) + k_{ir} \cdot \quad (4)$$

where $k_{ir}$ is the relative clock drift between the receiver and the *i*th station (receiver minus station) which is assumed to be constant from $t_i$ to $t_j$.

Based on (3) and (4), (2) is rewritten as follows.

$$\rho_j(t_j) = D_j(t_j) + b_i(t_i) + k_{ir} \cdot \quad b_{ij}(t_j) + d_{itx} + d_{rx} + \varepsilon(t_j) \quad (5)$$

In this way, all pseudorange measurements from different base stations and at different time instants are connected to the measurement from the *i*th station at $t_i$.

Other variables in (5) including $d_{itx}$ and $d_{rx}$ can be modeled as constants, and therefore can be measured beforehand. There are still too many unknowns in (5) which should be further simplified. Back to (1), we can observe that the clock offset between the *i*th and *j*th base stations, i.e. $b_{ij}$ in (5), can be possibly estimated if the transmission and reception time, the distance and the transmit and receive delays are obtained.

Furthermore, the user device position is embedded in the receiver and base station distance term $D$, as given by (6).

$$D_j(t_j) = \sqrt{(x_j - x(t_j))^2 + (y_j - y(t_j))^2 + (z_j - z(t_j))^2} \quad (6)$$
$$= \sqrt{(x_j - x(t_i) - v_x \cdot (t_j - t_i))^2 + (y_j - y(t_i) - v_y \cdot (t_j - t_i))^2 + (z_j - z(t_i) - v_z \cdot (t_j - t_i))^2}$$

where $x$, $y$ and $z$ are the respective 3-axis coordinates of the user position, $v_x$, $v_y$ and $v_z$ are the 3-axis velocities. The velocity of the user is assumed constant during this period from $t_i$ to $t_j$.

Therefore, the user position and velocity at any other time instant are connected with the ones at $t_i$. As a result, the unknowns to be resolved are the receiver 3-axis position and velocity at $t_i$, and the clock offset and drift at the same instant between the receiver and the *i*th base station.

III. SYSTEM DESIGN

The localization system is comprised of base stations and user devices or receivers as shown in Fig. 1. When in operation, the base stations broadcast positioning signals and listen to other base stations, and the receivers receive and process the positioning signals from several base stations to compute their own positions.

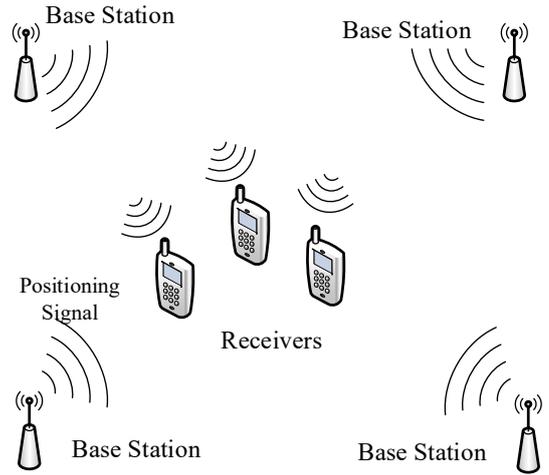

Fig. 1. Positioning System Constituent

A. Base Station

Based on (1), in order to obtain the clock bias between two base stations, the transmitting base station should put its send time information into the broadcasting signal and the receiving base station needs to measure the reception time based on its own clock ticks. Since the clocks of the base stations are drifting when time grows, the communication between base stations must be sufficiently frequent to keep up with the drifting.

Specifically, individual time slots are assigned to every base stations for signal transmission. During one time slot, the transmitting base station calculates its own clock offsets relative to other base stations at the pre-defined transmit time and puts them along with the transmit time into the payload of the positioning signal. During other time slots, the above station receives signals from other base stations and updates the relative clock offset information with them. The diagram of time slots and the tasks of base stations are shown in Fig. 2. It should be noted that the sent clock offsets should be aligned with the sending time to compensate the drift from its actual reception time. This can be done using a simple extrapolation or a filter[6].

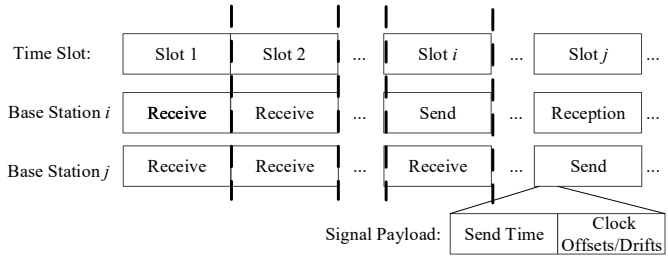

Fig. 2. Time sequence of base stations

### B. Receiver

A receiver receives the signal from one base station and measures its time of arrival to form a pseudorange measurement as given by (2). After several measurements are obtained, e.g. 6 measurements, an equation set can be established. According to the analysis in Section II, there are six unknowns to solve, i.e. the receiver 3-axis position and velocity and the clock offset and drift as given by (7).

$$X = \left[ x, y, z, b_i, v_x, v_y, v_z, k_{ir} \right]^T \quad (7)$$

Various methods can be used to solve this equation set to give the final solution of $X$. For example, the well-known conventional iterative least squares approach is definitely an option and is direct and simple for implementation[7].

It should be noted that the unknowns are the values at a specific time instant. Theoretically, the time instant $t_i$ can be selected arbitrarily. However, the time instant when the received information contains the maximal number of clock offsets with other base stations can be selected to maximize the number of equations and finally to obtain a possibly more accurate position solution. Besides, other strategies such as outlier identification and rejection, smoothing and filtering of the results should be considered and adopted.

## IV. TEST AND RESULT

### A. System Implementation

A chip that sends and receives ultra-wide band signals is adopted both in base stations and receivers to avoid the cumbersome debugging and testing work on radio frequency circuits and thus expedite the manufacturing of the hardware[8]. A micro-controller unit is used to control the chip and also complete the positioning computation in the receiver. Wifi modules are installed on the receivers and the base stations to transfer their data back to a central computer for monitoring. The hardware block diagrams of a base station and a receiver are depicted in Fig. 3.

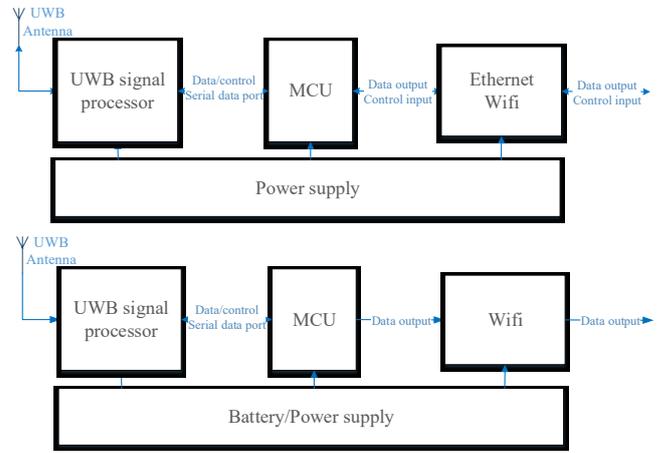

Fig. 3. Base station (top) and receiver (bottom) hardware

### B. Test Settings

Six base stations are used to cover an area of about 31 m long and 10 m wide. The base stations are placed at the sides and corners of the area and their coordinates are given in Fig. 4. The receivers are put on a turntable inside this area as indicated in the same figure.

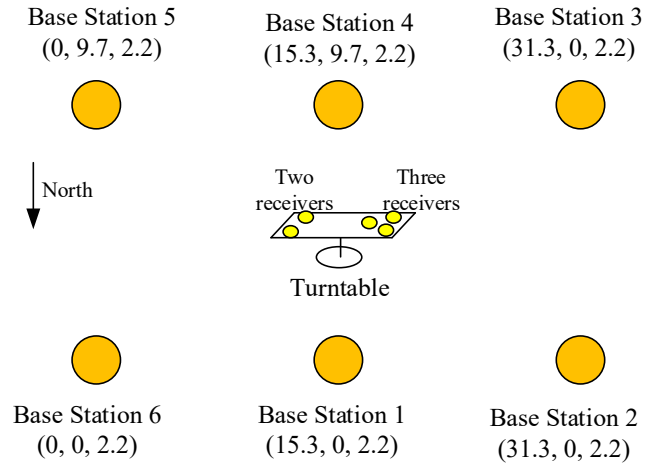

Fig. 4. Base station and receiver placement

Five receivers are respectively placed on the two arms of the turntable, 3 on one arm and 2 on another. The photograph of the placement during the test is given in Fig. 5.

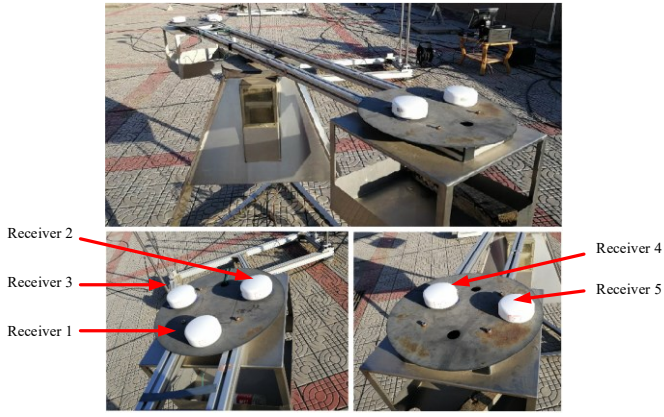

Fig. 5. Receiver placement on the turn table. Top: full view of the five receivers on the turntable. Bottom left: 3 receivers on one arm. Bottom right: 2 receivers on another arm.

Static and dynamic tests are conducted. In the static test, all the receivers are stationary on the turntable and their positioning results are transferred to the central computer and recorded. In the dynamic test, the receivers are also placed at the same place on the turntable as in the static test. The turntable is doing a circular motion with a constant linear velocity and the positioning results of the receivers are recorded for analysis.

*C. Result and Discussion*

The static positioning results of the entire area in the horizontal plane are shown in Fig. 6, and the magnified region of the five receivers is shown in Fig. 7. It can be seen that the relative positions of all the receivers are correctly computed and are identical with that in the photograph in Fig. 5.

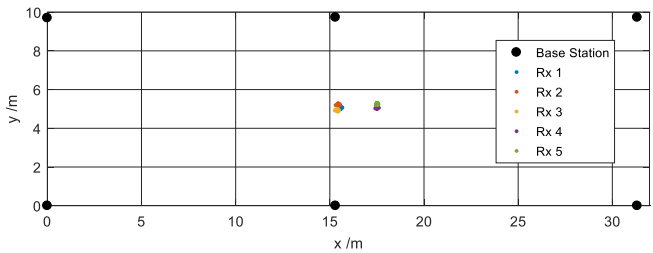

Fig. 6. Static positioning results

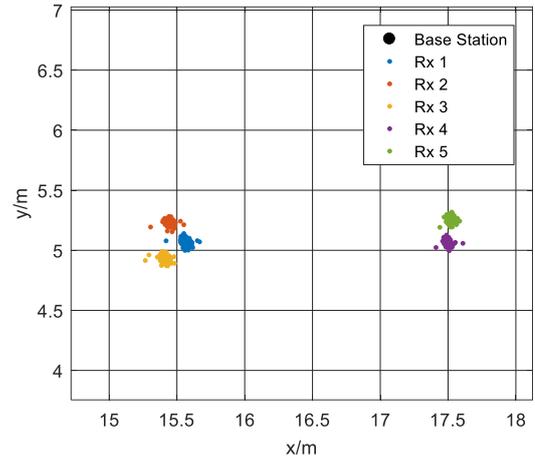

Fig. 7. Static positioning results (magnified)

The x and y curves are shown in Fig. 8. The standard deviation of x and y results are given in TABLE I. This result validates the positioning algorithm in the static condition.

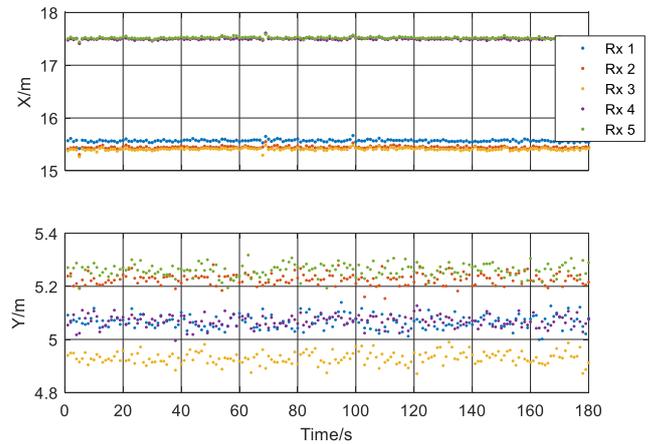

Fig. 8. Static positioning curves

TABLE I. Standard deviation of x and y position results

| Receiver No. | x /m | y /m |
|---|---|---|
| 1 | 0.0230 | 0.0253 |
| 2 | 0.0227 | 0.0209 |
| 3 | 0.0235 | 0.0228 |
| 4 | 0.0173 | 0.0222 |
| 5 | 0.0177 | 0.0237 |

The dynamic results are depicted in Fig. 9. The circular trajectories of all the receivers prove the correctness of the positioning results.

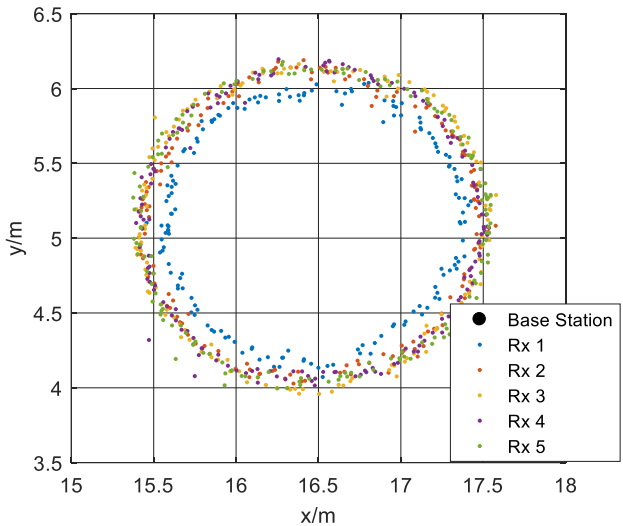

Fig. 9. Dynamic positioning results (magnified)

The sine wave curves demonstrated in Fig. 10 shows the individual positioning result for every receiver. It can be seen that the actual uniform circular motion is correctly reflected by the position solutions.

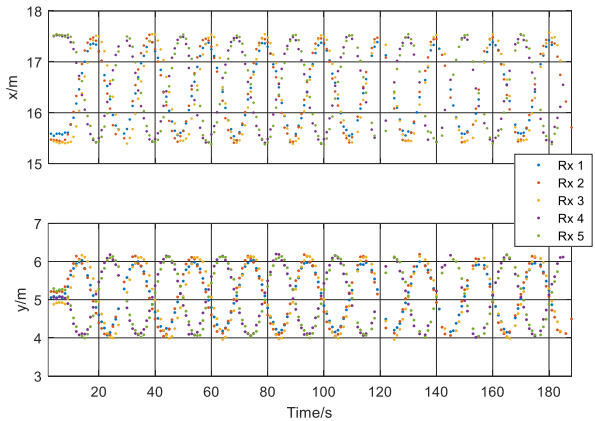

Fig. 10. Dynamic positioning curves

Baselines formed by two receivers are also investigated to test the relative accuracy of the position solutions. The baseline length results of receiver 1 and 4 as well as receiver 2 and 5 are given by Fig. 11. The standard deviations of the two baselines are 4.5 cm and 4.9 cm, respectively, as shown in the figure.

Both static and dynamic test results indicate that the receivers obtains the position solutions correctly. The positioning accuracy can be within 10-centimeter level. The position bias or offset compared with the true position is not considered yet because it is a determined value rather than a random value which can be compensated. For example, a cumbersome but at least feasible approach is that we can measure the interested area beforehand and compute the offset between the measured coordinates and the ones generated by the receivers and then apply this offset to other receiver-generated position results.

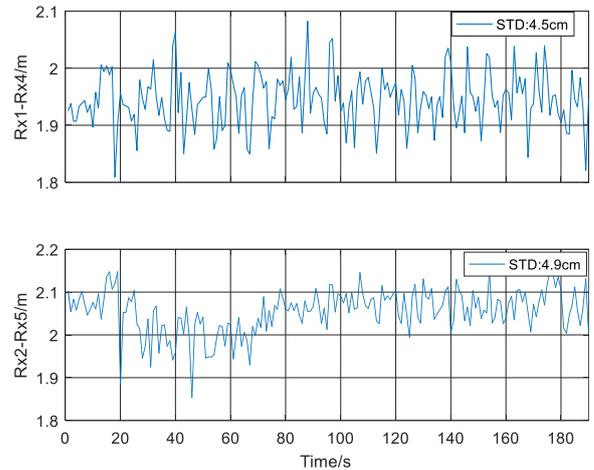

Fig. 11. Dynamic baseline results

## V. ACKNOWLEDGMENT

The authors are grateful to Mr. Zhengshen Xiao for his help on setting up the test environment.

## VI. CONCLUSION AND FUTURE WORK

A positioning system in which user devices receives signals from base stations to obtain their position results is proposed. The mathematical model of the system is proposed and analyzed based on which the broadcasting base stations and the receivers are designed and implemented. Field tests are conducted to validate the correctness of proposed theory and the performance of the system. Both static and dynamic results show that the positioning results have an accuracy better than 10 centimeters.